\documentclass{PoS}
\usepackage{amsmath}
\usepackage{gensymb}
\usepackage{subcaption}
\usepackage{url}

\title{Towards a Luminosity Function of TeV Gamma-ray Blazars}

\ShortTitle{Towards a Luminosity Function of TeV Gamma-ray Blazars}

\author{\speaker{Aryeh Brill}, for the VERITAS Collaboration\footnote{\protect\url{https://veritas.sao.arizona.edu/}} \footnote{for collaboration list see PoS(ICRC2019)1177}\\
        Columbia University, Department of Physics, New York, NY, USA\\
        E-mail: \email{aryeh.brill@columbia.edu}}

\abstract{Over seventy blazars identified as sources of TeV gamma ray emission have been detected, including approximately sixty BL Lac objects and seven flat spectrum radio quasars. The distribution in space of these objects can be described by a luminosity function, which gives their comoving space density as a function of their luminosity. We investigate a source selection method to be used for determining the luminosity function of TeV gamma-ray blazars using observations from VERITAS, a ground-based gamma-ray observatory consisting of an array of four atmospheric Cherenkov telescopes located in southern Arizona.}

\FullConference{36th International Cosmic Ray Conference -ICRC2019-\\
		July 24th - August 1st, 2019\\
		Madison, WI, U.S.A.}

\begin{document}

\section{Introduction}

\begin{figure}
\centering
\begin{subfigure}{.5\textwidth}
  \centering
  \includegraphics[width=0.9\linewidth]{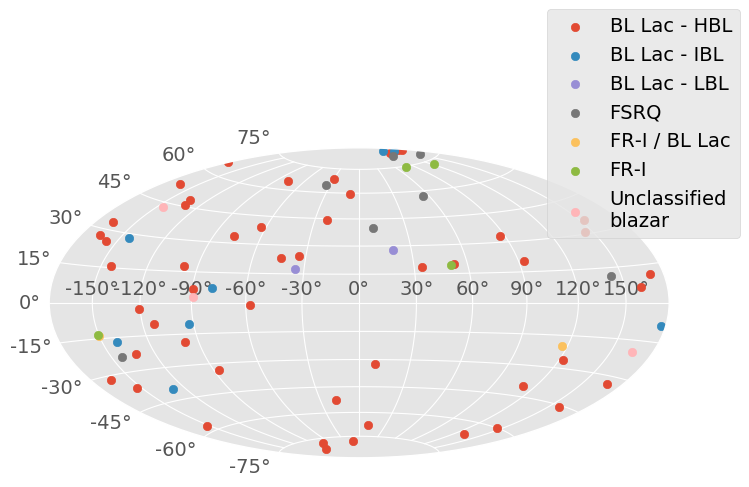}
  \label{fig:skymap}
\end{subfigure}%
\begin{subfigure}{.5\textwidth}
  \centering
  \includegraphics[width=0.9\linewidth]{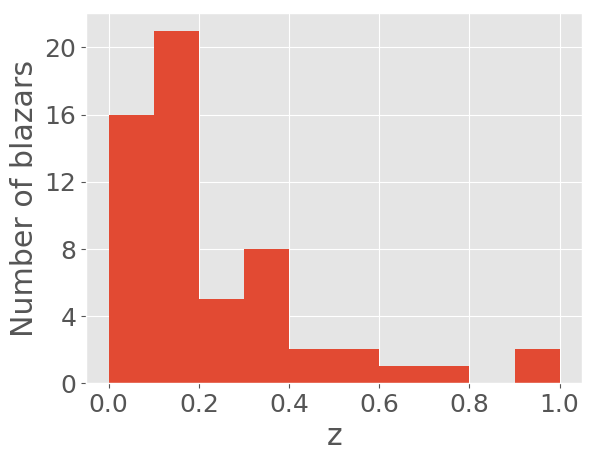}
  \label{fig:redshifts}
\end{subfigure}
\caption{Left: The extragalactic TeV sky in Galactic coordinates \cite{TeVCat}. Right: Redshift distribution of the 60 TeV blazars with known redshifts.}
\label{fig:tev_sources}
\end{figure}

Imaging atmospheric Cherenkov telescopes (IACTs) have detected over seventy TeV gamma-ray blazars, including about sixty BL Lac objects and seven flat spectrum radio quasars (FSRQs; Figure~\ref{fig:tev_sources}). These detections come from over a decade of observations by both current-generation IACTs and the \textit{Fermi} gamma-ray space telescope. With this rich dataset, we can study TeV blazars not just as individual sources but as a population. A key to understanding TeV blazars as a population is deriving their luminosity function (LF). The LF gives the comoving space density of these sources as a function of their luminosity, describing how they are distributed in space.

Estimating the LF of TeV blazars has numerous scientific motivations. To start, integrating the LF and subtracting the contribution from resolved sources provides an estimate of the contribution of blazars to the diffuse extragalactic gamma-ray background (EGRB) at $\sim$100 GeV and above. Blazar jets may produce neutrinos as well as gamma rays via pion production if they contain relativistic protons. Extrapolating the energy density of the gamma-ray blazar EGRB contribution to that of the isotropic high-energy neutrino flux using a power-law emission model provides a source-independent way to estimate the extent to which blazars produce high-energy neutrinos \cite{Ahlers2018}. In addition, emission above $\sim$50-100 GeV can produce election-positron pairs and generate secondary gamma rays via cascade emission, contributing to the MeV-GeV EGRB \cite{Coppi1997}. With the TeV blazar LF, this contribution can be computed and used to place constraints on unknown populations and exotic physics.

Additionally, fitting an evolution model to the LF describes blazar population change over cosmic time. Comparing how different blazar classes and objects observed at other wavelengths evolve can indicate whether they belong to the same fundamental population. Finally, fitting a model of Doppler boosted luminosities to the observed LF lets us recover the intrinsic LF and distribution of source Lorentz factors \cite{Urry1991}, informing our understanding of the parent population of these sources and of jet physics in blazars.

The LF of gamma-ray blazars has been previously calculated at energies below the TeV band, most notably using the FSRQs and BL Lac objects detected by \textit{Fermi}-LAT at 100 MeV - 100 GeV \cite{Ajello2012} \cite{Ajello2013}, as well as earlier using the smaller sample of blazars detected by EGRET between 20 MeV and 30 GeV \cite{Chiang1998}. Here, a preliminary investigation is conducted to study the requirements for deriving the luminosity function of TeV blazars using archival data from VERITAS, which introduces several complications not present with data from all-sky blazar surveys conducted at lower energies. VERITAS is an array of four IACTs sensitive to gamma rays with energies between 100 GeV and >30 TeV, located in southern Arizona. It has a 3.5\degree field of view and observes only under clear, dark skies. VERITAS regularly monitors known and candidate TeV blazars and follows up flaring events seen in its own and multiwavelength observations \cite{Benbow2017}.

\section{Challenges for Creating a TeV Blazar Luminosity Function}

A common approach for determining the LF and evolution of a population of sources given a flux-limited survey is to use a maximum likelihood (ML) estimator to fit a parameterized model using each source's luminosity, redshift, and any other source properties to be modeled. This method avoids binning the data, thereby reducing information loss and preventing potential biases caused by evolution within the bins, at the cost of introducing the assumption that the LF take a particular functional form.

However, performing a population study with a pointed instrument like VERITAS is more complicated than with all-sky instruments. The short, targeted observations performed by IACTs are highly sensitive to blazars' significant flux variability. Observations may be taken disproportionately during flaring states, biasing estimates of average flux. The effects of absorption on the extragalactic background light (EBL), too, become important at very high energies. IACTs have an overall low integration time and their sky coverage depends strongly on observation strategy, which is generally not geared toward conducting a uniform survey. Among other biases, IACT observations are not randomly distributed on the sky, but concentrated on known sources (either in gamma rays or other wave bands). Given this bias, it is not valid to consider IACT observations as a uniform survey limited by the sensitivity of TeV observations alone. The remainder of this article considers a method to counteract source selection bias, leaving other challenges for future work.

\section{Source Selection Study}

Constructing a valid TeV blazar LF requires emulating a uniform, flux-limited survey. One method to do this is to consider subsets of targets in a multiwavelength catalog that fulfill physically motivated selection criteria. For example, a \textit{Fermi}-LAT blazar catalog could be used, or an X-ray one motivated by the production of TeV emission by X-rays undergoing inverse Compton scattering. This procedure matches how VERITAS chooses bright sources from catalogs of blazars of various source classes for blazar discovery observations \cite{Benbow2017}, e.g. 2WHSP, an infrared catalog of high synchrotron peaked blazars \cite{Chang2017}, for HBLs. The selection criteria and catalog sensitivity are then used to derive an approximate TeV flux limit. Assuming all TeV emitters of the given source class are in the underlying catalog, and that the catalog itself provides an unbiased, complete sample, this process results in a complete, uniform survey up to the TeV flux limit. The chosen catalog and selection criteria must maximize completeness while minimizing false positives. Of course, in order to use archival data, the selected sources that are visible must already have been observed. Multiple catalogs may be combined to obtain sources of different classes.

The above method can be illustrated using the objects selected by Costamante \& Ghisellini (2002; CG02) \cite{Costamante2002}, who produced a list of 33 candidate (and 5 already-known) TeV BL Lac objects using selection limits on the X-ray and radio energy flux. Their sources were selected from a set of BL Lac samples for which radio, optical and X-ray observations were all available, including some surveys with coverage of the entire Northern sky, with others looking at bright sources only. Of the 38 candidate and known sources, 30 now have TeV detections \cite{TeVCat} and all have 4FGL associations \cite{FermiLAT2019}. 31 are visible to VERITAS (declination between $-10\degree$ and $+70\degree$), which has detected 20 and published upper limits on 8 of them \cite{Archambault2016}.

\begin{figure}
    \centering
    \includegraphics[width=0.7\linewidth]{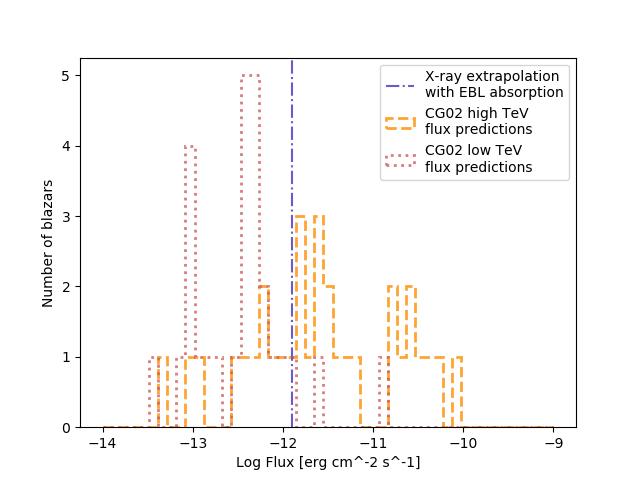}
    \caption{Distribution of the predicted TeV fluxes above 0.3 TeV from Costamante \& Ghisellini (2002; CG02) \cite{Costamante2002} from EBL absorption applied using Model C of \cite{Finke2010}. EBL absorption was applied using the redshift of each source from TeVCat \cite{TeVCat} or SIMBAD \cite{Wenger2000}, or $z=0.2$ when not available, and average energy assuming a power-law spectrum with $E_{th} = 0.3~\text{TeV}$ and $\Gamma = 3.5$. The "low" and "high" flux predictions refer to the SSC model and parameterization of Fossati et al. (1998) \cite{Fossati1998} from CG02. The dashed-dotted blue line is the CG02 candidate selection X-ray flux limit $F_X = 1.46~\mu\text{Jy}$ at $\nu_X = 1~\text{keV}$, extrapolated to the TeV band using the relation of \cite{Stecker1996} with EBL absorption applied as above.}
    \label{fig:flux_state}
\end{figure}

\begin{figure}
    \centering
    \includegraphics[width=0.7\linewidth]{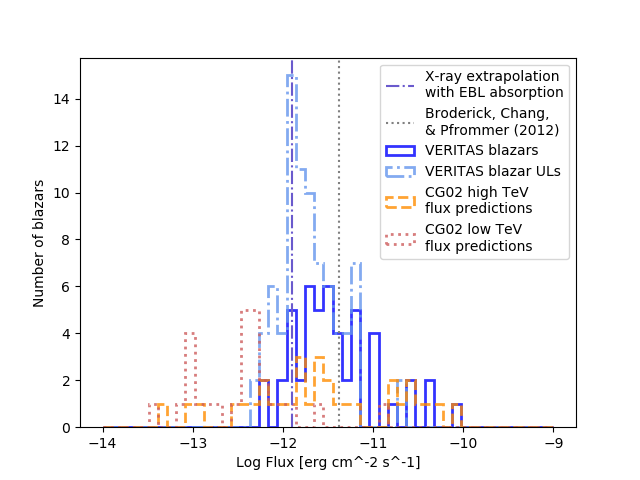}
    \caption{Distribution of the fluxes of VERITAS detected blazars and upper limits, overlaid on the CG02 flux predictions shown in Figure~\ref{fig:flux_state}. The solid dark blue histogram shows the energy fluxes of the 33 BL Lac objects detected by VERITAS, using fluxes in Crab Units and spectral indices from TeVCat \cite{TeVCat}, assuming $F_{\text{Mrk 501}} = 0.85~\text{Crab}$ and $\Gamma = 3.5$ for any source with no reported index. The energy fluxes were calculated by converting the Crab fluxes into photon fluxes above 0.3 TeV, assuming a Crab power-law spectrum with index 2.49 and normalization $3.2\times10^{-11}~\text{cm}^{-2}~\text{s}^{-1}~\text{TeV}^{-1}$. The photon flux was converted into energy flux above 0.3 TeV using the spectral index, assuming a power law spectrum. No attempt was made to distinguish average fluxes or spectra from those in flaring or other states. The dashed-dotted light blue histogram shows the distribution of VERITAS blazar upper limits from \cite{Archambault2016}, with differential flux limits converted energy flux assuming $E_{\text{th}} = 0.3~\text{TeV}$ and $\Gamma = 3.5$ for all sources. Also shown is the empirical flux limit of Broderick, Chang, \& Pfrommer (2012), $4.19\times10^{-12}~\text{erg}~\text{cm}^{-2}~\text{s}^{-1}$ \cite{Broderick2012}.}
    \label{fig:L_relation}
\end{figure}

\begin{figure}
    \centering
    \includegraphics[width=0.7\linewidth]{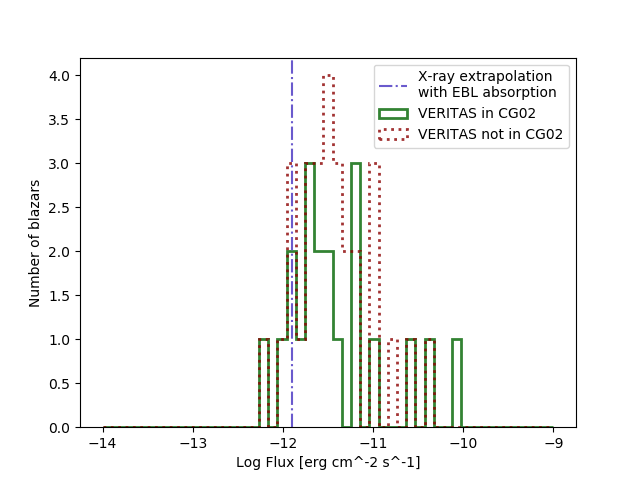}
    \caption{Distribution of energy fluxes of BL Lac objects detected by VERITAS from Figure~\ref{fig:L_relation}, split into those included in the candidates or known sources of Costamante \& Ghisellini (2002; CG02) \cite{Costamante2002} (solid green), and those that are not (dotted red).}
    \label{fig:incompleteness}
\end{figure}

The CG02 TeV candidate BL Lac objects can be used to explore the potential, and possible pitfalls, of this source selection approach for emulating a TeV blazar survey. To be useful, a source selection method based on an external catalog must both allow for the establishment of an effective TeV flux limit and provide a reasonably complete sample. The results of a preliminary investigation using these sources as the selected sample are shown in Figures~\ref{fig:flux_state},~\ref{fig:L_relation},~and~\ref{fig:incompleteness}.

\section{Discussion}

One way to obtain a predicted TeV flux given the observed fluxes at other wavelengths is to derive it from a model of the spectral energy distribution (SED) fitted to the multiwavelength data. Fortunately, this is exactly what CG02 have already done. They apply two models, a one-zone Synchrotron Self-Compton (SSC) model and the parameterization of \cite{Fossati1998}, built to describe sources with synchrotron and self-Compton peaks of equal power, and use both to predict the energy fluxes above 40 GeV, 0.3 TeV and 1 TeV without incorporating absorption by the EBL. Figure~\ref{fig:flux_state} shows the distributions of the flux predictions above 0.3 TeV for the two models, with EBL absorption using the Model C from Finke et al. (2010) \cite{Finke2010} additionally applied.

A second way to obtain a TeV flux limit is to use a simple physically motivated relationship between luminosities at different wavelengths, such as that of Stecker et al. (1996), $\nu_{TeV}F_{TeV} \sim \nu_{X}F_{X}$ for X-ray selected BL Lac objects \cite{Stecker1996}. An extrapolation using this relationship of the X-ray selection flux limit of CG02, $F_X~=~1.46$~$\mu$Jy, with the same EBL absorption correction applied, is also shown in Figure~\ref{fig:flux_state}. The predicted fluxes span over three orders of magnitude, with the extrapolated X-ray limit over an order of magnitude higher in flux than the lowest TeV prediction of each of the two models. For these predictions and luminosity relation to provide a useful flux limit, the predicted fluxes should cut off sharply at the low-flux end at a level consistent with the extrapolated flux limit. However, this behavior is not evident.

In addition, to take the selected sources as the basis for a flux-limited sample, the sensitivity of actual observations must match the supposed flux limit. Figure~\ref{fig:L_relation} shows the predicted fluxes of CG02 overlaid with the fluxes of the BL Lac objects actually detected by VERITAS, as well as upper limits of blazar discovery targets observed but not detected. The reported blazar fluxes plotted in Figure~\ref{fig:L_relation} do not necessarily come from similar emission states, limiting the physical interpretation of this distribution, but  are here intended for characterizing the sensitivity of VERITAS to these sources. A rough drop-off in both the detected and constrained fluxes is apparent around $1\times10^{-12}~\text{erg}~\text{cm}^{-2}~\text{s}^{-1}$. This value approximately matches the extrapolated X-ray flux limit, but is significantly higher than the lowest fluxes predicted by CG02 using spectral modeling.

Also shown is the empirical flux limit of Broderick, Chang, \& Pfrommer (2012) at  $4.19\times10^{-12}~\text{erg}~\text{cm}^{-2}~\text{s}^{-1}$, which was derived from a sample of 28 objects with publicly available well-defined SEDs observed by H.E.S.S., MAGIC, and VERITAS \cite{Broderick2012}. This limit appears too high to describe well the flux distributions from VERITAS or the predictions of CG02, indicating that a careful consideration of the sample being used is necessary when defining a TeV flux limit. 

Finally, for the source selection method to be useful, not only should VERITAS have observed all of the selected sources, but the converse must also be true: the source selection must be complete in the sense that all of the sources detected by VERITAS above the effective TeV flux limit are included. In fact, this is not the case. Of the 33 BL Lac objects detected by VERITAS, only 20 are CG02 known sources or candidates, and the other 13 are not in the catalog, a $\sim 40\%$ incompleteness rate. Figure~\ref{fig:incompleteness} shows the distributions of these two subpopulations. Visually, the distributions do not differ substantially, particularly at the critical low-flux end, showing that this incompleteness cannot be captured by a simple difference in flux levels (such as excluded sources being dimmer).

Although the study performed here does not allow a definite conclusion to be drawn, a possible explanation for both the lack of a clear limiting flux and the sample incompleteness is suggested. Blazars, including BL Lac objects, are highly variable across the electromagnetic spectrum. This variability matters in the context of source selection in several ways.

First, sources in a low state when measured by multiwavelength surveys could fall below the selection criteria flux level but still be TeV emitters. The X-ray, optical, and radio data available in the literature to CG02 to set selection cutoffs and assemble SEDs were not necessarily simultaneous, which, in addition to uncertainties from their model choices and parameters, likely played a role in incomplete source selection. For example, one VERITAS source missed by CG02, W~Comae, was identified as a promising candidate for TeV emission by a study that performed detailed modeling of its simultaneous broadband SED and X-ray variability \cite{Boettcher2002}. Since simultaneous multiwavelength measurements are not guaranteed to exist in the literature for every blazar, large uncertainties in predicted fluxes and selection thresholds are to some extent inherent in any selection method for TeV blazar candidates relying on archival data.

In addition, the difficulty of predicting TeV fluxes reflects not only uncertainties in extrapolating from lower wavelengths, reducible with simultaneous measurements and detailed modeling, but also actual variability in the TeV emission. TeV blazar detectability thus depends both on the limiting flux and the flux state when observed. A source might only be detected if it by chance flared while being observed. While over many sources this effect would average out without changing population characteristics, for smaller samples it may present a source of systematic uncertainty.

\section{Conclusion}

Determining the luminosity function of TeV-emitting blazars is an important step for understanding these sources as a population. Over a decade of gamma-ray data on these sources now exists. Over 70 TeV blazars have now been detected, and the upcoming Cherenkov Telescope Array (CTA), with an order of magnitude increase in sensitivity, should detect many more. Although the LF of gamma-ray blazars has been determined using data from all-sky high-energy surveys, extending these methods for a
population of very-high-energy TeV blazars poses new challenges: correcting for selection effects, incorporating variability, and accounting for EBL absorption.

This work presents a preliminary study of a method to use targeted observations to emulate a uniform survey by restricting the sources considered to those satisfying multiwavelength selection criteria. Setting an effective flux limit and obtaining a complete sample are both found to be challenging. The variability of blazars at all wavelengths can explain these difficulties.

\section{Acknowledgements}

This research is supported by grants from the U.S. Department of Energy Office of Science, the U.S. National Science Foundation and the Smithsonian Institution, and by NSERC in Canada. This research used resources provided by the Open Science Grid, which is supported by the National Science Foundation and the U.S. Department of Energy's Office of Science, and resources of the National Energy Research Scientific Computing Center (NERSC), a U.S. Department of Energy Office of Science User Facility operated under Contract No. DE-AC02-05CH11231. We acknowledge the excellent work of the technical support staff at the Fred Lawrence Whipple Observatory and at the collaborating institutions in the construction and operation of the instrument.

\end{document}